\documentstyle[preprint,prd,aps]{revtex}
\begin{document}

\title{Pair Production of charged vector bosons in \\ supercritical magnetic
fields at finite temperatures}

\author{V.R. Khalilov$^1$ and Choon-Lin Ho$^2$}

\address{\small \sl
1. Department of Physics, Moscow State University\\
Moscow $119899$, Russia\\
2. Department of Physics, Tamkang University, Tamsui 25137, Taiwan}

\maketitle

\begin{abstract}
The thermodynamic properties of an ideal gas of charged vector bosons
(with mass $m$ and charge $e$)
is studied in a strong external homogeneous magnetic field no greater than the
critical value $B_{cr}=m^2/e$.
The thermodynamic potential, after appropriate analytic continuation, is then
used in the study of the spontaneous
production of charged spin-one boson pairs  from vacuum in the presence of
a supercritical homogeneous magnetic field at finite temperature.

\end{abstract}

\vskip 1 cm
\pacs{13.40.-f, 05.30.-d, 05.30.Jp}

\newpage

\section{Introduction}

As is well known \cite{NOS}  the  energy spectrum of the vector
boson
with mass $m$, charge $e$, spin $S=1$ and gyromagnetic ratio $g=2$ in a
constant uniform magnetic field
${\bf B}=(0,0,B)$ is given by the formula (we set
$c=\hbar=1$)
\begin{eqnarray}
 E_n(p)=\sqrt{m^2+(2n+1-2S)|eB|+p^2},\qquad S=-1,\ 0,\ 1
\label{e1}
\end{eqnarray}
The integer $n$ ($n=0,1,2,\ldots$) labels the Landau level, and
$p$ is the momentum
along the direction of the field . For $n=0,\ p=0,\ S=1,\  E_0$ vanishes at
$B=B_{\rm cr}\equiv m^2/e$.
When $B>B_{\rm cr},\ E_0$ becomes purely imaginary.
Such a behavior of the energy
$E$ reflects a quantum instability of electrically charged vector
boson field in the presence  of an external uniform  magnetic field. The
source
of this instability is due to the interaction of
the external field with the additional
(anomalous) magnetic moment of the bosons, which, owing to the
gyromagnetic ratio $g=2$ ,
appears already in the tree approximation.

Charged spin-1 particles with the  gyromagnetic
ratio $g=2$ are not minimally coupled to an external electromagnetic
field ( if they were coupled in such a way the above ratio had to be
$g=1$).
However, the quantum theory of relativistic charged spin-1 bosons with
$g=2$ in the presence of external electromagnetic fields is a linear
approximation of gauge field theory in which the local $SU(2)$ symmetry
is spontaneously broken into $U(1)$ symmetry \cite{HS}.
So one anticipates that the perturbative vacuum of
the Weinberg--Salam electroweak model in the linear approximation would
exhibit
instability in a homogeneous superstrong magnetic field.

When $B$ becomes equal to $B_{\rm cr}$ the lowest energy levels
of charged spin-1 particle and
antiparticle ``collide'' with each other at $E_0=0$.  One finds similar
behavior in the case of scalar particles in a deep potential well  which
acts as the external field\cite{Mig}.  In this latter case, one usually
interprets
the behavior of energies
as follows: when the binding energy of a state exceeds the
threshold for particle creation, pairs of scalar particle-antiparticle may be
spontaneously produced giving rise to the so-called condensate.
The number of boson pairs produced by such a supercritical external field
(here the depth of the well) may be limited if only the mutual interaction of
the created particles
is taken into account. In the framework of (second) quantized field theory the
behavior of bosons in supercritical external fields was first considered
in \cite {Mig1}, which assumes a self-interaction of the
$\phi^4$-type, with the conclusion that the vacuum is, in fact,
stabilized by the
extremely strong (mutual) vacuum polarization.  For a thorough
discussion on the problem of electron-positron and scalar boson pair
production in external electromagnetic fields see \cite{RFK}.

The case of vector bosons was considered in \cite{OPK} by
taking into account only the ground state of the charged
spin-1 bosons in the superstrong external magnetic field and assuming a
self-interaction for this state like the one of the $W_{\mu}$ vector boson
field in the Weinberg-Salam electroweak gauge theory, namely the $|W|^4$
interaction.
In this work the condensate energy of charged spin-1 boson pairs was found,
and
a scheme for quantizing the $W$ field in the neighborhood of the new classical
vacuum with $W_{\rm classical}\ne 0$ near the threshold for condensate
production
$B-B_{\rm cr}\ll B_{\rm cr}\equiv m^2_W/e$ ($m_W$ is the mass of the $W$
boson) was presented.

Using the complete electroweak Lagrangian
the authors of~\cite{ambj}
have managed to construct new ``classical'' static magnetic solutions for a
$W$-condensate in the tree approximation.  They also show that
the instability of the $W$ field does not occur owing to the $|W|^4$
self-interaction term in the electroweak Lagrangian.  Moreover,
the electroweak gauge symmetry may
be restored in the presence of superstrong magnetic field with
$B=m^2_H/e$ ($m_H$ is the mass of the Higgs boson) if $m_H>m_W$.

In the one-loop approximation of the effective Lagrangian of the
charged spin-1 boson field (without a self-interaction term), radiative
corrections may induce,
in the presence of a strong uniform magnetic fields, the production of charged
vector boson
pairs in the lowest energy states, {\it i.e.} the condensate.  It is of
interest
to see what happens with the vacuum when not only an external magnetic field
is present but also when the temperature is finite.

In this paper we shall investigate the problem of pair production of charged
vector bosons induced by the unstable mode
in the presence of a supercritical magnetic field at finite temperature.
To study the vacuum effects we need to compute the effective potential
density, which is closely related to the
thermodynamic potential.  To this end we shall first try to treat the problem
in the framework of standard quantum statistical physics for
the case $B<B_{\rm cr}$ when quantum statistical quantities such as the
thermodynamic potential are unambiguous and may be well defined.
After deriving the thermodynamic potential in the region $B<B_{\rm cr}$ we
shall perform an analytic
continuation of this quantity into the supercritical region $B>B_{\rm cr}$.
This will give us the imaginary part of the effective potential, from which
we can derive the expression for the rate of pair production.
The contribution in the thermal one-loop effective action  from
gauge boson field in a constant homogeneous magnetic field was previously
considered
in \cite{chak} in connection with the question of symmetry restoration.
But in that work contribution from the unstable modes was
explicitly ignored.

\section{Thermodynamic potential}

The thermodynamic potential $\Omega$ for a gas of real (not virtual)
charged vector bosons as a function of the chemical potential $\mu$, the
magnetic induction of external field $B$, and the gas temperature
$T\equiv 1/\beta$ is defined by
\begin{eqnarray}
\Omega= {eBV\over 4\pi^2\beta}~\int dp~\ln\left\{1-\exp\beta\left[
\mu-\left(m^2-eB+p^2\right)^{1/2}\right]\right\}\phantom{mmmmmmmmmmmm}
\nonumber \\ + {eBV\over
4\pi^2\beta}~\sum^\infty_{n=0}~g_n~\int dp~\ln\left\{1-\exp\beta\left[
\mu-\left(m^2+(2n+1)eB+p^2\right)^{1/2}\right]\right\}~,
\label{Omega}
\end{eqnarray}
where $V$ is the volume of the gas, and $g_n=3-\delta_{0n}$ counts the
degeneracy of the excited states.

By expanding the logarithms and integrating over $p$, one can recast $\Omega$
into \cite{KHY}
\begin{eqnarray}
\Omega(\mu)= \Omega_1 +\Omega_2 +\Omega_3 \phantom{mmmmmmmmmmm}\nonumber \\
=-{VeB \over 2\pi^2\beta}
~\sum^\infty_{k=1}k^{-1}\exp(k\beta\mu)\left[M_{-}K_1(k\beta M_-)~
-M_{+}K_1(k\beta M_+)\right]~
\nonumber \\
-{3VeB \over 2\pi^2\beta}~\sum^\infty_{n=0}
\sqrt{M_{+}^2+2neB}
~\sum^\infty_{k=1}k^{-1}\exp(k\beta\mu)K_1\left(k\beta~
\sqrt{M_{+}^2+2neB}\right)~,
\label{Omega2}
\end{eqnarray}
where $M_{\mp} \equiv \sqrt{m^2 \mp eB}$ and $K_n (x)$ is the modified Bessel
function
of order $n$. If both particles and antiparticles are present,
the factor $\exp(k\beta\mu)$ in (\ref{Omega2}) has to be replaced by
$2\cosh(k\beta\mu)$.
The thermodynamic
potential as a function of the
chemical potential is real-valued for real values of $\mu$ that are to satisfy
for particles and antiparticles with  mass $M_-$ the inequality $|\mu|\le
M_{-}$. This condition comes from the physical requirement that the density
(and the occupation numbers) of particles and antiparticles with the mass
$M_-$ is positive for any real values of momenta $p$.

In weak field $B\ll m^2/e$ and at temperatures~
$eB/m \ll T < m$ when the spacing between Landau levels is still considerably
less than the thermal energy, one can approximate $\Omega$ as follows.
For $\Omega_1$ and $\Omega_2$ in (\ref{Omega2}), we
set $M_{\mp}\approx m(1 \mp \chi/2)$ with $\chi\equiv eB/m^2$, and use the
following formula
\begin{eqnarray}
 K_1(k\beta m(1 \mp \chi/2)) = K_1(k\beta m) + \chi \frac{dz}{d\chi}
\frac{dK_1(z)}{dz}~,
\label{expand}
\end{eqnarray}
where $z= k\beta m(1 \mp \chi/2)$.  Evaluation of $\Omega_3$ can be done by
first replacing the summation over
$n$ by an integral using the Euler formula
\begin{eqnarray}
 \sum^\infty_{n=0}f(n+1/2) = \int\limits_0^\infty f(x)dx + (1/24)f'(0)~,
\label{Eul}
\end{eqnarray}
 and then by using the formula \cite{GR}
\begin{eqnarray}
\int^\infty_1 dz~z^2~K_1(kmz\beta)
=\frac{1}{km\beta}K_2(km\beta)~.
\label{K}
\end{eqnarray}
The thermodynamic potential and density of spin-1 bosons at equilibrium
can then be obtained as
\begin{eqnarray}
\Omega &\simeq& -{VT^{1/2}m^{3/2}\over (2\pi)^{3/2}}\left[3T^2{\rm Li}_{5/2}
(e^{\beta(\mu-m)}) +\frac{7(eB)^2}{8m^4}e^{\beta(\mu-m)}\right]~,
\label{Omega4}\\
\rho&\simeq& 3\left({Tm\over 2\pi}\right)^{3/2}~\zeta(3/2)~,
\label{rho1}
\end{eqnarray}
where ${\rm Li}_s (x) =\sum_{k=1}^\infty x^k/k^s$ is the
polylogarithmic function of order $s$, and ${\rm Li}_s (1)=\zeta(s)$.
The magnetization of the gas under the above conditions is
a positive function\footnote{we take this opportunity to correct the
expression for the magnetization in \cite{KHY}.} of the magnetic
field induction and temperature because paramagnetic (spin) contribution
dominate
\begin{eqnarray}
 M_z(B)= -\frac{1}{V}\frac{\partial \Omega}{\partial B} =
\frac{7e^2BT^{1/2}}{4(2\pi)^{3/2}m^{1/2}}e^{\beta(\mu-m)}.
\label{M-rho+}
\end{eqnarray}

When $B\approx B_{\rm cr}$
transitions of bosons from level $n=0$ to any excited levels
$n\ge1$ will not be allowed if $T<eB/m$ and all bosons in quantum state
with $n=0$ that are available may be considered as condensate in
a two-dimensional ``momentum''
space in the plane perpendicular to the magnetic field with values
of ``effective momenta'' $k<(eB)^{1/2}$.  True condensate, however, will not
actually be formed in three-dimensional momentum space because longitudinal
momenta of bosons may have values outside this region.

For low temperature $T$ such that $\beta M_-\gg 1$, contributions to the
thermodynamic potential (\ref{Omega2}) from all the excited states of the
vector bosons are exponentially small compared with that from the state
with $n=0$ and $S=1$.
Hence only the first term $\Omega_1$ in (\ref{Omega2}) needs be considered
in this limit
\begin{eqnarray}
 \Omega(\mu) = \Omega_1 (\mu)
=-{VeBM_-\over 2\pi^2\beta}
 \sum^\infty_{k=1}k^{-1}\exp(k\beta\mu)K_1(k\beta M_-)~.
\label{Omega2b}
\end{eqnarray}
Subsequently, the boson density is
\begin{eqnarray}
 \rho_g=\frac{eBM_-}{2\pi^2}
 \sum^\infty_{k=1}K_1(k\beta M_-) \exp\left(k\beta \mu\right)~.
\label{e7}
\end{eqnarray}
When $M_{-}\gg T$,  $M_{-}-\mu<T$, we can get for
the total density at equilibrium
\begin{eqnarray}
 \rho\cong \frac{eB(TM_{-})^{1/2}}{(2\pi)^{3/2}}\left[\left(\frac{\pi T}
{M_{-}-\mu}\right)^{1/2}-1.46\right].
\label{eq22}
\end{eqnarray}
The first (leading) term of (\ref{eq22}) reduces exactly to the
one obtained in \cite{Rojas}.
The total boson density (\ref{e7}) for relatively ``high'' temperature for
which $T>M_{-}$ but $T\ll m$ is
\begin{eqnarray}
 \rho\cong \frac{eBT}{2\pi^{2}}e^{\mu/T},
\label{eq23}
\end{eqnarray}
for $-\mu\gg T$
and
\begin{eqnarray}
 \rho\cong \frac{eBT}{2\pi^{2}}\ln(T/|M_{-}+\mu|)
\label{eq23a}
\end{eqnarray}
for $-\mu\to M_-$.

It follows from formulae (\ref{eq22}) and (\ref{eq23a}) that
significant amount of vector bosons persists to fill the states with non-zero
momentum projections on the magnetic field direction.  These states now
should be considered as excited ones. Hence, any density of bosons can be
accommodated outside the ground state (with $p=0$) at any temperature.  This
means that there is no true Bose--Einstein condensation in the presence of
finite magnetic fields.   We mention here that it was R. Feynman \cite{Fnm}
who first showed that true BEC is impossible in a classical
one-dimensional gas.

An exact expression for the magnetization of the vector boson gas in the
lowest energy state in a strong magnetic field may be derived from
(\ref{Omega2b}) in the form \cite{KHY}
\begin{eqnarray}
  M_z(B)&=&{e\over 2\pi^2\beta}
\sum_{k=1}^{\infty}\left[\frac{M_-}{k}
K_1(k\beta M_-)
+{eB\beta\over 2} K_0(k\beta M_-) \right]
\exp(k\beta\mu)~. \label{M-}
\end{eqnarray}
The magnetization is also a positive function of the magnetic
field and temperature.

\section{Pairs production of vector bosons}

Let us now turn to discussing the problem of pair production of vector bosons
in a supercritical magnetic field at finite temperature.  There are two
possible mechanisms for this process:
1) pairs may be produced as a result of thermal collisions of real charged
bosons in the external field,
2) pairs  may be  spontaneously produced by a constant magnetic  field
when $B>B_{\rm cr}$ from the vacuum, just as
electron-positron pairs are produced by an external
electric field \cite{schw}.

Before we proceed to the second mechanism, let
us first give some estimates of the density of spin-1 boson pairs (in the
lowest energy state only) that may be produced as a result of
thermal collisions of real bosons in the external field with
$B\approx B_{\rm cr}$.
If the density of the created pairs is much greater than
that of the bosons present initially, we may apply formula~(\ref{e7}) with
$\mu=0$ to find the density of the pairs produced by thermal
collisions.  For low ($\beta M_->1$ but $T\ll m$) and ``high''
($\beta M_- <1$, $T<m$) temperature, we obtain respectively
\begin{eqnarray}
\rho_{T} \cong \frac{eB(M_- T)^{1/2}}{(2\pi)^{3/2}}\exp(-M_-/T),
\label{eq14}
\end{eqnarray}

\begin{eqnarray}
\rho_{T} \cong \frac{eBT}{2\pi^2}\ln(T/M_-).
\label{eq15}
\end{eqnarray}

Now we come to the second mechanism.
As is known (see, for example \cite{GMM}) the quantum electrodynamics vacuum
in the presence of an external electromagnetic field can be described by the
total transition amplitude from the vacuum state $|0_{\rm in}\rangle$
in time $t\to -\infty$ to the vacuum state $\langle 0_{\rm out}|$ in
time $t\to \infty$ as follows:
\begin{equation}
\label{ampl}
 C_{\rm v} = \langle 0_{\rm out}|0_{\rm in}\rangle = \exp(iW({\cal E}, B))~,
\end{equation}
where $W$ is the effective action for a given quantum field.
$W$ defines the effective Lagrangian $L_{\rm eff}$ according to $W=\int d^4 x
L_{\rm eff}$.  The effective action is a classical functional
depending on the external electric ($\cal E$) and magnetic ($B$) fields.
When the external electromagnetic field is homogeneous
the effective action is equal to
$W({\cal E}, B) = -(E({\cal E}, B)-E({\cal E}=0,B=0))V\Delta t$, where
$E({\cal E},B)$ is nothing but  the density of vacuum energy in the presence
of the external field, $V$ is the volume and
$\Delta t$ is the transition time. It is worthwhile to note that the
effective action contains all divergencies of the theory but they are in the
real part of $W({\cal E}, B)$. $C_{\rm v}$ is the probability amplitude
when the external electromagnetic field does not change and so this applies
for the vacuum.

For external fields smoothly changing both in space and time one has
\begin{equation}
 |C_{\rm v}|^2 = \exp(2\Im L_{\rm eff}({\cal E}, B)V\Delta t)~.
\label{Cv}
\end{equation}
The imaginary part of the effective Lagrangian density $\Im L_{\rm eff}$, or
of the vacuum energy,
is finite and describes production of particles
by the external electromagnetic field.
It also signals that an instability of the vacuum occurs.
The imaginary part of the effective Lagrangian density  reduces at $T=0$
to the imaginary part of the effective potential density.  The latter
(for the case under consideration) arises from the lowest energy
of the charged massive vector boson
being imaginary at $B>B_{\rm cr}$.
At $T=0$ the imaginary part of the effective potential is
\cite{NOS,Khal}:
\begin{equation}
\Im V_0=-\frac{eB(eB-m^2)}{16\pi}.
\label{seventeen}
\end{equation}
Here $V_0$ is the zero-point energy density of the vacuum at $T=0$.
>From (\ref{Cv}) one concludes that the quantity $\Gamma (B)\equiv -2\Im
V_0(B)$ is the production rate
per unit volume of vector boson pairs (or, equivalently, the decay rate of
the vacuum) in the external magnetic field at $T=0$, which, in this case, is
given by \begin{equation}
\Gamma (B) =  \frac{eB(eB-m^2)}{8\pi}.
\label{Gamma0}
\end{equation}

When $T\neq 0$ one must use the thermal one-loop effective potential density
which is related to the thermodynamic potential by
(see, for example \cite{chak})
\begin{equation}
\label{eff}
 V(B, T) = V_0+ \Omega (\mu=0,B,T)/V~.
\end{equation}
The first term in (\ref{eff})
does  not depend  on temperature, while the second term coincides with the
thermodynamic potential (per unit volume) of noninteracting gas of bosons  at
$\mu=0$.
We see that the thermodynamic potential at $\mu=0$  play the role of the free
energy of the vacuum of quantum field system in the presence of the
external field  at finite temperature.

Since the  unstable
mode contributes to $\Omega_1$, it follows from (\ref{Omega2b}) that
the temperature-dependent part of the effective potential
also becomes complex when $B>B_{\rm cr}$.
The point $B=B_{\rm cr}$ is the branch point of the effective potential
considered as a function of product $eB$.
One could  avoid the complex values of the effective potential by taking
account
of the ``physical'' part of energy spectrum (\ref{e1}) which  may be
well defined only when $B<B_{\rm cr}$.

To find the effective potential in the region $B>B_{\rm cr}$ we must perform
an analytic
continuation of $K_1(z)$ as a function of variable $z=k\beta M_-$ into  the
complex range.  Let us  denote  the
argument of function $K_1(z)$ in the region $B>B_{\rm cr}$ as~
$z=\pm ik\beta \sqrt{eB-m^2} \equiv\pm ikt$.
Then, by Schwartz symmetry principle, we have
\begin{equation}
\label{Schw}
K_1(\pm ikt)=K_1^*(\mp ikt)=-\frac{\pi}{2}\left[ J_1(kt)\mp
iY_1(kt)\right]~,
\end{equation}
where $J_1(kt)$ and $Y_1(kt)$ are the first- and
second-order  Bessel functions, respectively.
>From (\ref{Omega2b}) and (\ref{Schw}) we get
\begin{equation}
\label{twone}
\Omega_1 (B,T) =
\frac{Vm^3}{4\pi\beta}\chi\sqrt{\chi-1}\sum\limits_{k=1}^\infty
\frac{1}{k}\left[\pm iJ_1(k\beta m\sqrt{\chi-1})+Y_1(k\beta
m\sqrt{\chi-1})\right]~.
\end{equation}
Here $\chi=eB/m^2$ as defined before.
For the  imaginary part of $\Omega_1 (B,T)$ one can obtain another form using
the following formula which is valid for small temperatures such that
$t>2\pi$ :
\begin{equation}
\label{twtwo}
\sum\limits_{k=1}^\infty\frac{J_1(kt)}{k}=1-\frac{t}{4}+\frac{2}{t}
\sum_{n=1}^{l}\sqrt{t^2-(2\pi n)^2},
\end{equation}
where $l$ is the integral part of $t/2\pi$.

The production rate at finite temperature and magnetic field is now
defined to be $\Gamma (B,T)\equiv -2\Im V (B,T)$ according to (\ref{Cv}).
Since $\Gamma (B,T)$ must be positive at
any finite temperature, $\Im V(B,T)$ must always be negative.  This
requirement allows one to perform the analytic
continuation  unambiguously. It follows from Eqs. (\ref{seventeen}),
(\ref{eff}), (\ref{twone}) and (\ref{twtwo}) that we must take
(\ref{twone}) with the lower sign in order to obtain a non-negative decay
rate:
\begin{equation}
 \Gamma (B,T) =  \frac{m^4}{8\pi} \chi\left(\chi-1\right)\left[1+
\frac{4}{x} \sum\limits_{k=1}^\infty \frac{J_1(kx)}{k}\right]~,
\label{prodpar}
\end{equation}
where $x\equiv \beta m\sqrt{\chi -1}$.  This is the total pair
production rate of vector bosons induced by the unstable mode in a
supercritical
magnetic field at finite temperatures.  We note here that the expression
inside
the square bracket of (\ref{prodpar}) depends on $\beta,m,e$ and $B$ only
through the combination $x$.

For low temperature ($\sqrt{eB-m^2}\gg T$)
 formula (\ref{prodpar}) may be simplified. Applying the asymptotic expansion
for $J_{\nu}(z)$ at $z\to\infty$ in the form
$$
 J_{\nu}(z) = \sqrt{2/\pi z}\cos(z-\nu\pi/2-\pi/4)~,
$$
replacing the summation over $k$ by the integration and then taking into
account the following formula for the Fourier integral \cite{olw}
\begin{eqnarray}
 \int\limits_a^\infty e^{ixt}f(t)dt = ie^{iax}f(a)/x + o(1/x) \quad
(x\to\infty)
\label{rim}
\end{eqnarray}
(by the Riemann-Lebesgue lemma the last formula is valid if the integral
converges uniformly in $(a, \infty)$ at all large enough $x$),  we obtain
(up to the term $o(1/x)$)
\begin{equation}
 \Gamma (B,T) \sim  \frac{m^4\chi (\chi-1)}{4\pi}\left[1 +
\left(\frac{2}{x\pi^{1/5}}\right)^{5/2}\cos\left(x-\frac{\pi}{4}
\right)\right]~.
\label{prodlow}
\end{equation}
As $T\to 0$ ($x\to \infty$), (\ref{prodlow}) reduces to the zero temperature
expression (\ref{Gamma0}).

\section{Resume}

Our calculations have been  performed for  the  case  of  the
constant (in time) magnetic  field.
When pairs  are  spontaneously produced by  the  constant magnetic  field
exclusively the background
magnetic field will  be  changed  in time. One may suppose that the background
magnetic field is likely   to   remain  constant  in  time  only   during   a
characteristic  time $\delta t\sim 1/\sqrt{eB-m^2}$.

We see that the number of boson pairs produced by such a supercritical
external field increases with increasing magnetic field. This number
may be limited if the mutual interactions of the created particles
are taken into account.   As mentioned in the Introduction the vacuum may be
stabilized
with the appearance of vector boson condensate (at $T=0$) in the tree
approximation.
Another possibility to stabilize the vacuum is the one-loop radiative
corrections to the mass of the charged spin-1 boson field in the critical
field region $B\to B_{\rm cr}$ \cite{klin}.

\section{Acknowledgments}

This work was supported in part by the Republic of China through Grant
No. NSC 88-2112-M-032-002.
VRK would like to thank the Department of Physics of Tamkang University
(R.O.C.) for kind hospitality and financial support.

\vfil\eject

\end{document}